\documentclass[12pt]{article}
\usepackage{amsmath}
\usepackage{cite}

\begin{document}

\begin{center}
{\Large \bf Lift of Invariant to Non-Invariant Solutions of
Complex Monge-Amp\`ere Equations}
\\[4mm]
{\large \bf M B Sheftel$^{1,2}$ and A A Malykh$^2$}
\\[3mm] $^1$ Department of Physics, Bo\u{g}azi\c{c}i University,
34342 Bebek, Istanbul, Turkey \\ $^2$ Department of Higher
Mathematics, North Western State Technical University, Millionnaya
St. 5, 191186, St. Petersburg, Russia \vspace{1mm}
\\ E-mail: mikhail.sheftel@boun.edu.tr , specarm@mail.wplus.net
\end{center}

\begin{abstract}
\noindent We show how partner symmetries of the elliptic and
hyperbolic complex  Monge-Amp\`ere equations ($CMA$ and $HCMA$)
provide a lift of non-invariant solutions of three- and
two-dimensional reduced equations, i.e., a lift of invariant
solutions of the original $CMA$ and $HCMA$ equations, to
non-invariant solutions of the latter four-dimensional equations.
The lift is applied to non-invariant solutions of the
two-dimensional Helmholtz equation to yield non-invariant
solutions of $CMA$, and to non-invariant solutions of
three-dimensional wave equation and three-dimensional hyperbolic
Boyer-Finley equation to yield non-invariant solutions of $HCMA$.
By using these solutions as metric potentials, it is possible to
construct four-dimensional Ricci-flat metrics of Euclidean and
ultra-hyperbolic signatures that have non-zero curvature tensors
and no Killing vectors.
\end{abstract}

\section{Introduction}

Solutions of Pleba\~nski's first and second heavenly equations
yield a potential that determines Ricci-flat (anti-)self-dual
metrics on $4$-dimensional complex manifolds \cite{pleb}. In other
words, these "heavenly" metrics satisfy complex vacuum Einstein
equations. In the case of the first heavenly equation, physically
important ones are two real cross sections of these complex
metrics, K\"ahler metrics with Euclidean or ultra-hyperbolic
signature, when the first heavenly equation coincides with the
elliptic ($CMA$) and hyperbolic ($HCMA$) complex Monge-Amp\`ere
equation respectively. In particular, among the solutions
$u(z_1,\bar z_1,z_2,\bar z_2)$ of the elliptic $CMA$
\begin{equation}
   u_{1\bar 1} u_{2\bar 2} - u_{1\bar 2} u_{2\bar 1} = 1
   \label{cma}
\end{equation}
there are gravitational instantons, the most important of which is
$K3$, the Kummer surface \cite{ahs}. The explicit construction of
the $K3$ metric is still an unsolved challenging problem. The main
difficulty is that this metric should have no Killing vectors and
therefore the corresponding solution of $CMA$ should be a
non-invariant solution (with no point symmetries). That was a
basic motive for us to develop methods for finding non-invariant
solutions of nonlinear partial differential equations (PDEs). In
the context of Lie's theory of symmetries of differential
equations, the standard method for solving nonlinear PDEs is
symmetry reduction that yields only invariant solutions and
therefore cannot produce K\"ahler potential of $K3$. Recently, we
have developed method of partner symmetries that yields
non-invariant solutions to the elliptic and hyperbolic $CMA$ and
to the second heavenly equation. Using them as metric potentials,
we have obtained some new heavenly metrics with no Killing vectors
\cite{mns,mnsgr,mnsh}.

Here we develop further our method of partner symmetries so, that
we are able now to obtain non-invariant solutions of the
four-dimensional $CMA$ and $HCMA$ starting from invariant
solutions of these equations that satisfy the corresponding
reduced equations of lower (three and two) dimensions and these
"seed" solutions should be non-invariant solutions of the reduced
equations. We have called this procedure {\it "lift"}. In
particular, we have obtained non-invariant solutions of elliptic
$CMA$ (\ref{cma}) by the lift from solutions of two-dimensional
Helmholtz equation and non-invariant solutions of hyperbolic
$HCMA$
\begin{equation}
u_{1\bar 1} u_{2\bar 2} - u_{1\bar 2} u_{2\bar 1} = -1
\label{hcma}
\end{equation}
by the lift from solutions of three-dimensional wave equation and
of the three-dimensional hyperbolic Boyer-Finley equation
\cite{bf}:
\begin{equation}
\psi_{z\bar z} = {\rm e}^{\psi_x}\psi_{xx},
 \label{bf}
\end{equation}

In section \ref{sec-partner-cma} we discuss partner symmetries of
the elliptic and hyperbolic complex Monge-Amp\`ere equations.

In section \ref{sec-helm} we obtain non-invariant solutions of the
elliptic $CMA$ by the lift from two-dimensional Helmholtz
equation.

In section \ref{sec-wave} we make the lift of solutions of
three-dimensional wave equation to non-invariant solutions of the
hyperbolic $HCMA$.

In section \ref{sec-legendre} we use Legendre transformation of
$HCMA$ and equations for rotational partner symmetries to obtain
hyperbolic Boyer-Finley equation ($BF$) and B\"acklund
transformations for $BF$ that we discovered earlier \cite{mnsw}.

In section \ref{sec-lift}, using results of the previous section,
we obtain non-invariant solutions of $HCMA$ by the lift from our
non-invariant solutions to hyperbolic $BF$ (\ref{bf}) that we
obtained earlier \cite{msw}. Noninvariant solutions to the
elliptic $BF$ were obtained first by D. Calderbank and P. Tod
\cite{CalTod}. A little later, we had independently obtained these
solutions to the elliptic $BF$ and also non-invariant solutions to
the hyperbolic $BF$ \cite{msw} by our version of the method of
group foliation \cite{ns}. We had also proved non-invariance of
all these solutions.

By using non-invariant solutions of $HCMA$ as metric potentials,
it is possible to construct new metrics with ultra-hyperbolic
signature that have no Killing vectors \cite{lift}. A
comprehensive survey of results on four-dimensional anti-self-dual
metrics with the ultra-hyperbolic (neutral) signature was given by
M. Dunajski and S. West in \cite{dunaj}.

We are working now on a modification of this method in order to
obtain non-invariant solutions of the elliptic $CMA$ (\ref{cma})
by the lift from solutions of the three-dimensional elliptic $BF$
\cite{bf}
\begin{equation}
\psi_{z\bar z} + {\rm e}^{\psi_x}\psi_{xx} = 0
 \label{ellBF}
\end{equation}
to non-invariant solutions of the elliptic $CMA$ (\ref{cma}) with
the final goal to obtain Ricci-flat heavenly metrics with
Euclidean signature that admit no Killing vectors.

\section{Partner symmetries of the complex Monge-Amp\`ere equations}
\setcounter{equation}{0}
 \label{sec-partner-cma}

  The hyperbolic and elliptic $CMA$ have the same set of symmetries
whose characteristics $\varphi$ satisfy the condition
\begin{equation}
u_{2\bar 2}\varphi_{1\bar 1} + u_{1\bar 1}\varphi_{2\bar 2} -
u_{2\bar 1}\varphi_{1\bar 2} - u_{1\bar 2}\varphi_{2\bar 1} = 0.
\label{desym}
\end{equation}
Define the operators
\begin{equation}
L_1 = \lambda (u_{1\bar 2}D_{\bar 1} - u_{1\bar 1}D_{\bar 2}),
\quad L_2 = \lambda (u_{2\bar 2}D_{\bar 1} - u_{2\bar 1}D_{\bar
2}),
  \label{L_i}
\end{equation}
where $D_i, D_{\bar i}$ are operators of total derivatives with
respect to $z_i, \bar z_i$ and $\lambda$ is a complex constant.
Then the symmetry condition (\ref{desym}) can be expressed as a
total divergence
\begin{equation}
D_1L_2\varphi = D_2L_1\varphi
 \label{divsym}
\end{equation}
so that locally there exists a symmetry potential $\psi$ defined
by the differential equations
\begin{equation}
\psi_1 = L_1\varphi ,\quad \psi_2 = L_2\varphi .
 \label{psi}
\end{equation}
It is easy to see that if $\varphi$ satisfies (\ref{divsym}), then
$\psi$ also satisfies (\ref{divsym}) and so the potential $\psi$
of a symmetry $\varphi$ is itself a symmetry of $CMA$
\cite{mns,mnsh}. These $\varphi$ and $\psi$ are called {\it
partner symmetries}.

Differential equations (\ref{psi}) are recursion relations for
symmetries
\begin{equation}
  \psi = R_1\varphi,\quad \psi = R_2\varphi
  \label{recurs}
\end{equation}
with the recursion operators
\begin{equation}
  R_1 = D_1^{-1} L_1,\quad R_2 = D_2^{-1} L_2.
  \label{R}
\end{equation}
The transformation inverse to (\ref{psi}) is obtained by taking
complex conjugates of equations (\ref{psi}), solving them
algebraically with respect to $\varphi_1$ and $\varphi_2$, and
using $CMA$:
\begin{equation}
\varphi_1 = \mp\bar\lambda^{-1} (u_{1\bar 2}\psi_{\bar 1} -
u_{1\bar 1}\psi_{\bar 2}),\quad \varphi_2 = \mp\bar\lambda^{-1}
(u_{2\bar 2}\psi_{\bar 1} - u_{2\bar 1}\psi_{\bar 2})
 \label{phi}
\end{equation}
where the minus and plus signs correspond to the elliptic and
hyperbolic $CMA$ respectively. Note that if $|\lambda| = 1$ the
inverse transformation (\ref{phi}) reads
\begin{equation}
  \varphi = \mp R_1\psi,\quad \varphi = \mp R_2\psi,
  \label{inverse}
\end{equation}
and then for $HCMA$ there is a simplifying choice $\psi =
\varphi$, when the transformation (\ref{psi}) coincides with its
inverse (\ref{phi}) and becomes
\begin{equation}
\varphi_1 = \lambda (u_{1\bar 2}\varphi_{\bar 1} - u_{1\bar
1}\varphi_{\bar 2}),\quad \varphi_2 = \lambda (u_{2\bar
2}\varphi_{\bar 1} - u_{2\bar 1}\varphi_{\bar 2}).
 \label{ps=fi}
\end{equation}
For the elliptic $CMA$ the ansatz $\psi = \varphi$ implies the
trivial solution $\varphi = \psi = 0$.

We will also need the equations complex conjugate to (\ref{ps=fi})
\begin{equation}
\varphi_{\bar 1} = \lambda^{-1} (u_{2\bar 1}\varphi_{1} - u_{1\bar
1}\varphi_{2}),\quad \varphi_{\bar 2} = \lambda^{-1} (u_{2\bar
2}\varphi_{1} - u_{1\bar 2}\varphi_{2}).
 \label{barphi}
\end{equation}
If we choose for $\varphi$ a characteristic of any Lie point
symmetry of $HCMA$, then (\ref{ps=fi}) and (\ref{barphi}) become
differential constraints, joined to $HCMA$, that select some
particular solutions of $HCMA$. Equations (\ref{ps=fi}) and
(\ref{barphi}) are not independent: any three equations out of the
four ones imply the fourth equation together with $HCMA$ itself as
their algebraic consequences. Alternatively, any two of these four
equations together with $HCMA$ imply the remaining two
constraints. We shall use $HCMA$ together with the first equations
in (\ref{ps=fi}) and (\ref{barphi})
\begin{equation}
\varphi_1 = \lambda (u_{1\bar 2}\varphi_{\bar 1} - u_{1\bar
1}\varphi_{\bar 2}),\quad \varphi_{\bar 1} = \lambda^{-1}
(u_{2\bar 1}\varphi_{1} - u_{1\bar 1}\varphi_{2})
 \label{phi1b1}
\end{equation}
as basic independent equations: the original PDE and two
constraints for one unknown $u$.

\section{Lift of non-invariant solutions of elliptic\\ CMA
from two-dimensional Helmholtz\\ equation}
\setcounter{equation}{0}
 \label{sec-helm}

Introducing real variables $x, y, z, t$ by the relations $z_1 =
(x+iy)/2$, $z_2 = (t+iz)/2$, we consider the elliptic $CMA$
(\ref{cma}) in a real form
\begin{equation}
(u_{xx} + u_{yy})(u_{zz} + u_{tt}) - (u_{xt} + u_{yz})^2 - (u_{yt}
- u_{xz})^2 = 1 .
  \label{realcma}
\end{equation}
First, consider solutions that are invariant under translations in
$x$, selected by the condition $u_1 + u_{\bar 1} \equiv 2u_x = 0$,
for which (\ref{realcma}) reduces to
\begin{equation}
u_{yy}(u_{zz} + u_{tt}) - u_{yz}^2 - u_{yt}^2 =1 \iff
u_{yy}u_{2\bar 2} - u_{y2}u_{y\bar 2} = 1.
 \label{redcma}
\end{equation}
Applying to this the Legendre transformation $v = u - yu_y$, $p =
u_y$, we obtain the three-dimensional Laplace equation
\begin{equation}
v_{2\bar 2} + v_{pp} = 0.
 \label{Laplace}
\end{equation}
We consider now solutions of (\ref{Laplace}) that are invariant
under the symmetry generator $X = v\partial_v + \partial_p $ due
to the condition $v_p - v = 0$ and thus imply further reduction of
(\ref{Laplace}). Then $v = e^p\theta(z_2,\bar z_2)$, where
$\theta$ satisfies the Helmholtz equation
\begin{equation}
\theta_{2\bar 2} + \theta = 0.
 \label{helm}
\end{equation}

In order to make a lift from solutions of these low-dimensional
reduced linear equations to non-invariant solutions of
four-dimensional elliptic $CMA$, we need to arrive at these
equations without symmetry reduction. For this purpose we choose
the symmetries of translations and dilations
\begin{equation}
\varphi = u_1 + u_{\bar 1}, \qquad \psi = u - z_1u_1 - \bar
z_1u_{\bar 1},
 \label{transdil}
\end{equation}
respectively, for characteristics of partner symmetries in the
equations (\ref{phi}) (with the minus sign) and their complex
conjugates.

After the Legendre transformation
\begin{equation}
v = u - z_1u_1 - \bar z_1u_{\bar 1},\quad  p = u_1,\quad \bar p =
u_{\bar 1}
 \label{leg}
\end{equation}
in the new variables $p, \bar p, v$ formulas (\ref{transdil})
become $\varphi = p + \bar p,\quad \psi = v $ and elliptic $CMA$
(\ref{cma}) takes the form
\begin{equation}
v_{p\bar p}v_{2\bar 2} - v_{p\bar 2}v_{\bar p2} = v_{p\bar
p}v_{2\bar 2} - v_{p\bar p}^2
 \label{cmaleg}
\end{equation}
where $v = v(p,\bar p,z_2,\bar z_2)$. The Legendre transformation
(\ref{leg}) maps the resulting equations for partner symmetries
and the transformed $CMA$ (\ref{cmaleg}) to the following system
of five independent equations \cite{mns}
\begin{equation}
v_{pp} = A v_{p\bar p},\quad v_{p\bar 2} = C v_{p\bar p},\quad
v_{2\bar 2} = B v_{p\bar p}
 \label{veq}
\end{equation}
together with their complex conjugates. Here the coefficients are
defined as
 \[ A  = \frac{1 + v_p^2 + iv_2}{\Delta}, \quad
 B = \frac{v_2v_{\bar 2} + i(v_2 - v_{\bar 2})}{\Delta}, \quad
 C = \frac{v_pv_{\bar 2} + i(v_p - v_{\bar p})}{\Delta} \]
where $\Delta = 1 + v_pv_{\bar p}$. Equations (\ref{veq}) imply
one more equation \cite{mns}
\begin{equation}
v_{p\bar p} = 1 + v_pv_{\bar p}
 \label{6eq}
\end{equation}
as their differential consequence. We note once again that the
transformed $CMA$ (\ref{cmaleg}) is satisfied automatically on
solutions of (\ref{veq}) and (\ref{6eq}).

The logarithmic substitution $v = -\ln{w}$ linearizes equations
(\ref{veq}) and (\ref{6eq}) in the form
\begin{gather}
 w_{p\bar p} + w = 0, \qquad \qquad \qquad w_{pp} + w - iw_2 = 0,
\label{linv}
\\ w_{p\bar 2} - i(w_p - w_{\bar p}) = 0, \qquad
w_{2\bar 2} - i(w_2 - w_{\bar 2}) = 0 \nonumber
\end{gather}
plus two complex conjugate equations. System (\ref{linv}) implies
\begin{equation}
w_{2\bar 2} - (w_{pp} + w_{\bar p\bar p} - 2w_{p\bar p}) = 0.
 \label{lineq}
\end{equation}
If $p = \alpha + i\beta$, $\bar p = \alpha - i\beta$, then in real
variables $\alpha$ and $\beta$ (\ref{lineq}) becomes the
three-dimensional Laplace equation
\begin{equation}
w_{2\bar 2} + w_{\beta\beta} = 0
 \label{Laplace1}
\end{equation}
which coincides with (\ref{Laplace}) up to a change in variables,
while the first equation (\ref{linv}) coincides with the Helmholtz
equation (\ref{helm}) in the new variables.

Thus, if we know a solution of one of the two above-mentioned
equations that depends on arbitrary constants, we can consider
them as arbitrary functions of all other variables that do not
show up explicitly in this equation. Then all the other equations
(\ref{linv}) determine a dependence of these functions on these
parameters and so we obtain a non-invariant solution of the
transformed elliptic $CMA$ (\ref{cmaleg}) by lifting it from the
invariant solution that satisfies the reduced equation
(\ref{Laplace}) or (\ref{helm}).

For example, let us consider a lift from Helmholtz equation
(\ref{helm}), that coincides in new variables with the first
equation (\ref{linv}), starting from its solution of the form
\begin{gather}
 w = \sum_{j}^{}A_j(z_2,\bar z_2)\times \phantom{\left\{ e^{2s_j{\rm
Re}(\alpha_jp)} {\rm Re}\left(F_j e^{2i{\rm Im}(\alpha_jp)}+ e^{-2
s_j{\rm Re}(\alpha_jp)}{\rm Re}\right)\right\}}
 \label{solut}
  \\  \left\{ e^{2s_j{\rm
Re}(\alpha_jp)} {\rm Re}\left(F_j e^{2i{\rm Im}(\alpha_jp)}\right)
+ e^{-2 s_j{\rm Re}(\alpha_jp)} {\rm Re}\left(G_j e^{2i{\rm
Im}(\alpha_jp)} \right) \right\}.
 \notag
\end{gather}
Then other equations (\ref{linv}) imply the following restrictions
on solution (\ref{solut})
\[A_j = \exp{\left\{2\,{\rm Im}\Bigl(\bigl(\alpha_j^2(s_j^2 + 1) + 1\bigr)
z_2\Bigr)\right\}} ,\qquad s_j=\sqrt{1-1/|\alpha_j|^2},\]
 while $\alpha_j,F_j, G_j$ are arbitrary complex constants. This
 is a non-invariant solution of the Legendre-transformed $CMA$
 \cite{mns}.

\section{Lift of non-invariant solutions of hyperbolic CMA
from three-dimensional wave\\ equation}
 \setcounter{equation}{0}
 \label{sec-wave}

For the translational symmetry reduction we will need the real
form of $HCMA$ obtained by the change of variables $z_1 = (x +
iy)/2$, $\bar z_1 = (x - iy)/2$, $z_2 = (t + iz)/2$, $z_2 = (t -
iz)/2$:
\begin{equation}
(u_{xx} + u_{yy})(u_{zz} + u_{tt}) - (u_{xt} + u_{yz})^2 - (u_{yt}
- u_{xz})^2 = -1 .
  \label{realhcma}
\end{equation}
We consider solutions of (\ref{realhcma}), invariant under
translations in $x$, that are selected by the condition $u_x = 0$.
Then (\ref{realhcma}) reduces to
\begin{equation}
u_{yy}(u_{zz} + u_{tt}) - u_{yz}^2 - u_{yt}^2  = -1 .
  \label{redhcma}
\end{equation}
Applying to (\ref{redhcma}) the Legendre transformation
\begin{equation}
v = u - yu_y,\quad q = u_y
  \label{legendre1}
\end{equation}
we end up with the three-dimensional wave equation
\begin{equation}
v_{qq} = v_{tt} + v_{zz}
  \label{wave}
\end{equation}
for the new unknown $v = v(q,t,z)$.

Now we will not perform any symmetry reduction but use equations
(\ref{ps=fi}) and (\ref{barphi}) for partner symmetries with
$\varphi$ equal to the characteristic of the symmetry of
translations in $x$: $\varphi = u_1 + u_{\bar 1}$, and  $\lambda =
i$. With these choices, after the Legendre transformation similar
to (\ref{legendre1})
\begin{equation}
v = u - z_1u_{z1} - \bar z_1u_{\bar z_1},\quad p = u_x,\quad  q =
u_y
  \label{legendre2}
\end{equation}
the real form (\ref{realhcma}) of $HCMA$ becomes
\begin{equation}
(v_{pp} + v_{qq})(v_{tt} + v_{zz}) - (v_{pt} - v_{qz})^2 - (v_{pz}
+ v_{qt})^2 = v_{pp}v_{qq} - v_{pq}^2,
  \label{legcma}
\end{equation}
while (\ref{ps=fi}) and (\ref{barphi}) yield only three
independent equations \cite{mnsh}
\begin{subequations}
\label{partsym}
\begin{gather}
v_{qq} = v_{pz} + v_{qt},
 \label{partsyma}
 \end{gather}
 \begin{gather}
v_{pq} = v_{qz} - v_{pt},
 \label{partsymb}
\end{gather}
 \begin{gather}
v_{qq} = v_{tt} + v_{zz}.
 \label{partsymc}
  \end{gather}
\end{subequations}
Equation (\ref{partsymc}) formally coincides with the
three-dimensional wave equation (\ref{wave}) that determines
solutions of (\ref{legcma}), invariant under translations in $x$.
However, the unknown $v$ in this equation depends also on the
fourth variable $p$, so that (\ref{partsymc}) depends on an extra
parameter $p$ and therefore it is actually not the reduced
equation (\ref{wave}). Note that the transformed non-reduced
$HCMA$ (\ref{legcma}) is a consequence of the system
(\ref{partsyma}--\ref{partsymc}) and so the latter equations
determine some partial solutions of (\ref{legcma}).

The solution set of (\ref{partsymc}) can be written as the double
Fourier integral
\begin{gather}
v = \int\limits_{-\infty}^{+\infty}\int\limits_{-\infty}^{+\infty}
\left(a(p,\alpha,\beta)\exp{\left\{-i\left(\alpha t + \beta z +
\sqrt{\alpha^2
+ \beta^2}\,q\right)\right\}}\right. \nonumber \\
 \left. \mbox{} + b(p,\alpha,\beta)\exp{\left\{-i\left(\alpha t +
\beta z - \sqrt{\alpha^2 +
\beta^2}\,q\right)\right\}}\right)d\alpha\, d\beta.
 \label{fourier}
\end{gather}
Imposing the remaining equations (\ref{partsyma}) and
(\ref{partsymb}) on solution (\ref{fourier}), we finally obtain
the general solution of the system
(\ref{partsyma})--(\ref{partsymc})
\begin{gather}
v = \int\limits_{-\infty}^{+\infty}\int\limits_{-\infty}^{+\infty}
\left(a(\alpha,\beta)\exp{\left\{-i\sqrt{\alpha^2 +
\beta^2}\left(\frac{\sqrt{\alpha^2 + \beta^2 + \alpha}}{\beta} p + q\right)\right\}}\right. \nonumber \\
 \left. \mbox{} + b(\alpha,\beta) \exp{\left\{-i\sqrt{\alpha^2 +
\beta^2}\left(\frac{\sqrt{\alpha^2 + \beta^2 - \alpha}}{\beta} p -
q\right)\right\}}\right)
 e^{-i(\alpha t + \beta z)} d\alpha\, d\beta.
 \label{fouriersol}
\end{gather}
which is a partial solution of the Legendre-transformed $HCMA$
(\ref{legcma}).

Thus, using equations (\ref{partsyma})--(\ref{partsymc}), implied
by our choice of partner symmetries, we have made a lift of
solutions of the reduced equation (\ref{wave}) to a set of partial
solutions of four-dimensional Legendre-transformed $HCMA$
(\ref{legcma}).

\section{Rotational partner symmetries, Legendre transformation and Boyer-Finley equation}
\setcounter{equation}{0}
 \label{sec-legendre}

Boyer-Finley equation usually arises from rotational symmetry
reduction of $HCMA$, subjected to the combination of the point and
Legendre transformation in the first pair of variables $z_1, \bar
z_1$
\begin{equation}
  z_1 = e^{\zeta_1},\; \bar z_1 =
e^{\bar\zeta_1},\; \zeta_1 = \psi_q,\; \bar\zeta_1 = \psi_{\bar
q},\; u = q\psi_q + \bar q\psi_{\bar q} - \psi,\; u_{\zeta_1} =
q,\; u_{\bar\zeta_1} = \bar q .
 \label{legendre}
\end{equation}
Here we do not perform any symmetry reduction but still apply the
same transformation (\ref{legendre}) to $HCMA$ and the two
independent constraints (\ref{phi1b1}), rename $z_2 = z$ and
choose $\varphi$ as the rotational symmetry characteristic
\begin{equation}
  \varphi = i(z_1u_1 - \bar z_1u_{\bar 1}) = i(q - \bar q).
  \label{rot}
\end{equation}
Then $HCMA$ becomes
\begin{equation}
  \psi_{q\bar q}\psi_{z\bar z} - \psi_{q\bar z}\psi_{\bar qz} +
  e^{\psi_q + \psi_{\bar q}} (\psi_{qq}\psi_{\bar q\bar
  q} - \psi_{q\bar q}^2) = 0,
  \label{leghcma}
\end{equation}
where $\psi(q,\bar q,z,\bar z)$ is the new unknown, and the
constraints (\ref{phi1b1}) take the form
\begin{equation}
e^{\psi_{\bar q}}(\psi_{\bar q\bar q} + \psi_{q\bar q}) = \lambda
\psi_{q\bar z},\quad \lambda e^{\psi_{q}}(\psi_{qq} + \psi_{q\bar
q}) = \psi_{\bar qz} .
 \label{legrot}
\end{equation}
Now, we express $\psi_{q\bar z}$ and $\psi_{\bar qz}$ from
(\ref{legrot}) and substitute them into $HCMA$ (\ref{leghcma})
with the result
\begin{equation}
  \psi_{z\bar z} = e^{\psi_q + \psi_{\bar q}} (\psi_{qq} + 2\psi_{q\bar q} + \psi_{\bar q\bar
  q}).
  \label{BF}
\end{equation}
In the real coordinates $x, y$ in the complex $q$-plane ($q = x +
iy,\;\bar q = x - iy$), (\ref{BF}) becomes the (hyperbolic)
Boyer-Finley equation
\begin{equation}
\psi_{z\bar z} = e^{\psi_x}\psi_{xx}.
 \label{realBF}
\end{equation}
The constraints (\ref{legrot}) take the form
\begin{equation}
 \psi_{zx} + i\psi_{zy} = 2\lambda\left[e^{(\psi_x -
  i\psi_y)/2}\right]_x,\quad
  \psi_{\bar zx} - i\psi_{\bar zy} = 2\lambda^{-1}\left[e^{(\psi_x
  + i\psi_y)/2}\right]_x .
  \label{rotxy}
\end{equation}

The variable $y$ does not appear explicitly in the Boyer-Finley
equation (\ref{realBF}), and so it can be regarded as a parameter
of a symmetry group of this equation: a change of $y$ does not
affect the equation. Let $\omega$ be any symmetry characteristic
of the Boyer-Finley equation in the form
\begin{equation}
  \tilde{\psi}_{z\bar z} = \exp{(\tilde\psi_{xx})},
  \label{B_F}
\end{equation}
related to (\ref{realBF}) by the substitution $\psi =
\tilde\psi_x$. Then a symmetry characteristic of (\ref{realBF}) is
$i\omega_x$ (where the factor $i$ is introduced for convenience)
and the Lie equation for the group with the parameter $y$ reads
\begin{equation}
  \psi_y = i\omega_x .
  \label{Lie}
\end{equation}
Eliminating $\psi_y$ in the constraints (\ref{rotxy}) with the aid
of (\ref{Lie}) and then integrating the result with respect to
$x$, we obtain
\begin{equation}
  \omega_z = \psi_z - 2\lambda e^{(\psi_x + \omega_x)/2},\quad
  \omega_{\bar z} = -\psi_{\bar z} + 2\lambda^{-1} e^{(\psi_x -
  \omega_x)/2}.
  \label{Backlund}
\end{equation}
These are B\"{a}cklund transformations for the Boyer-Finley
equation that we discovered earlier \cite{mnsw}. The differential
compatibility condition $(\omega_z)_{\bar z} = (\omega_{\bar
z})_z$ of the system (\ref{Backlund}) reproduces the Boyer-Finley
equation (\ref{realBF}), while the compatibility condition in the
form $(\psi_z)_{\bar z} = (\psi_{\bar z})_z$ yields the equation
for symmetry characteristics of the Boyer-Finley equation
(\ref{B_F}):
\begin{equation}
  \omega_{z\bar z} - e^{\psi_x} \omega_{xx} = 0 .
  \label{omeg}
\end{equation}

Thus, without any symmetry reduction, the Boyer-Finley equation
arises as a linear combination of the Legendre-transformed $HCMA$
and differential constraints (\ref{legrot}) implied by our choice
of rotational symmetry for both partner symmetries. Furthermore,
the differential constraints themselves turn out to be
B\"{a}cklund transformations for the Boyer-Finley equation in a
new disguise.

\section{Lift of solutions of Boyer-Finley equation
to non-invariant solutions of {\mathversion{bold} $HCMA$}}
 \setcounter{equation}{0}
 \label{sec-lift}

Now consider a reverse procedure. We start with the
three-dimensional Boyer-Finley equation together with its
B\"{a}cklund transformations, differentiated with respect to $x$,
and consider a symmetry group parameter as the fourth coordinate
$y$ in these equations, according to (\ref{Lie}). Then we arrive
at $4$--dimensional Legendre-transformed $HCMA$ as a linear
combination of these three equations. As a consequence, the
partner symmetries lift three-dimensional non-invariant solutions
of $BF$, that are still invariant solutions of $HCMA$, to
four-dimensional non-invariant solutions of $HCMA$.

We start with non-invariant solutions to the hyperbolic $BF$
\begin{equation}
  v_{z\bar z} = (e^v)_{xx}
  \label{b_f}
\end{equation}
that we had obtained earlier in \cite{msw} by our version of the
method of group foliation. Those solutions involve a couple of
arbitrary holomorphic and anti-holomorphic functions $b(z)$ and
$\bar b(\bar z)$ that arise as "constants" of integrations. In our
construction, $BF$ equation (\ref{realBF}) and its solutions
depend also on the fourth variable, the parameter $y$, and hence
the "constants" of integration, $b$ and $\bar b$, should also
depend on $y$:
\begin{equation}
  v(x,y,z,\bar z) = \ln{[x+b(z,y)]} + \ln{[x+\bar b(\bar z,y)]} -
2\ln{(z+\bar z)}.
 \label{sol_v}
\end{equation}
$BF$ equations (\ref{realBF}) and (\ref{b_f}) are related to each
other by the substitution $v = \psi_x$ and hence solutions of
(\ref{realBF}), $\psi = \int v dx$, are obtained by integrating
the formula (\ref{sol_v}) with respect to $x$ with the "constant"
of integration $F(z,\bar z,y)$:
\begin{gather}
\psi = [x+b(z,y)]\ln{[x+b(z,y)]} + [x+\bar b(\bar z,y)]\ln{[x+\bar
b(\bar z,y)]} \nonumber \\
\mbox{} - 2x[\ln{(z+\bar z)} + 1] + F(z,\bar z,y).
 \label{sol_psi}
 \end{gather}
For arbitrary functions $b$, $\bar b$, and $F$, this is a solution
to $BF$ (\ref{realBF}). The unknown $y$-dependence in
(\ref{sol_psi}) is determined by the requirement that $\psi$
should also satisfy the Legendre-transformed $HCMA$
(\ref{leghcma}).

We substitute the expression (\ref{sol_psi}) for $\psi$ in $HCMA$
(\ref{leghcma}) and, since all the $x$-dependence in
(\ref{sol_psi}) is known explicitly, (\ref{leghcma}) splits into
several equations, corresponding to groups of terms with a
different dependence on $x$. We were able to solve these equations
and make a complete analysis of all possible solutions. They have
the form
\begin{subequations}
\label{sol}
\begin{gather}
\psi = [q + b(z)]\ln{[q + b(z)]} + [\bar q + \bar b(\bar
z)]\ln{[\bar q + \bar b(\bar z)]}\nonumber \\
- (q + \bar q)[\ln{(z + \bar z)} + 1] + \int\!\!\int \frac{b(z) +
\bar b(\bar z)}{(z + \bar z)^2}\,d z d\bar z + r(y),
 \label{sola}
 \end{gather}
\begin{gather}
\psi = [q + b(z)]\ln{[q + b(z)]} + [\bar q + \bar b(\bar
z)]\ln{[\bar q + \bar b(\bar z)]} \nonumber \\
- (q + \bar q)[\ln{(z + \bar z)} + 1] + \int\!\!\int \frac{b(z) +
\bar b(\bar z)}{(z + \bar z)^2}\,d z d\bar z \nonumber\samepage
\\ \mbox{} + 2iy\ln{\left(\frac{\bar z}{z}\right)} + r(y),
 \label{solb}
\end{gather}
\begin{gather}
\psi = [q + b(z)]\ln{[q + b(z)]} + [\bar q + \bar b(\bar
z)]\ln{[\bar q + \bar b(\bar z)]} \nonumber \\
- (q + \bar q)[\ln{(z + \bar z)} + 1] + \int\!\!\int \frac{b(z) +
\bar b(\bar z)}{(z + \bar z)^2}\,d z d\bar z \nonumber \\
+ 2i\int\ln{\left[\frac{\bar z + 2ik(y)}{z - 2ik(y)} \right]}\, d
y + r(y).
 \label{solc}
\end{gather}
\end{subequations}
Here $r(y)$ and $k(y)$ are arbitrary smooth real-valued functions,
while $b(z)$ and $\bar b(\bar z)$ are arbitrary holomorphic and
anti-holomorphic functions of one complex variable that arise when
the $y$-dependence of $b(z,y)$ and $\bar b(\bar z,y)$ is
completely determined. Solution (\ref{solb}) is a particular
simple case of the more general solution (\ref{solc}) when
$k(y)=0$.

Note that, by construction, we have obtained the solutions of
$HCMA$ that satisfy only one additional differential constraint,
the Boyer-Finley equation, though we have two constraint equations
produced by partner symmetries. If we require that both
constraints should be satisfied, we obtain a subset of solutions
that are invariant with respect to non-local symmetries of $HCMA$,
though this does not mean invariant solutions in the usual sense
\cite{mns,mnsh}. For solutions with such special property we have
\begin{gather}
 r(y) = 2(\alpha - \pi)y + r_0
 \label{r(y)1} \\
 r(y) = 2\alpha y + r_0
  \label{r(y)23}
\end{gather}
in (\ref{sola}) and $(\ref{solb}, \ref{solc})$ respectively. Here
$r_0$ is an arbitrary real constant and $\lambda = e^{i\alpha}$.

It can be proved that if the functions $b(z)$, $\bar b(\bar z)$
are not constants, the formulas (\ref{sola})--(\ref{solc}) yield
non-invariant solutions of (\ref{leghcma}). As a consequence, by
the reasoning similar to \cite{mnsh}, the ultra-hyperbolic metrics
governed by the potentials $\psi$ in (\ref{sola})--(\ref{solc})
have no Killing vectors \cite{lift}.

\section{Conclusions}

We are interested in obtaining non-invariant solutions of
four-dimensional heavenly equations because they will yield new
gravitational metrics with no Killing vectors. This is a
characteristic property of the famous gravitational instanton $K3$
where the metric potential should be a non-invariant solution of
the elliptic complex Monge-Amp\`ere equation. Constructing an
explicit metric on $K3$ is our final goal. In this paper we have
used a new approach for solving such a problem which we call
"lift". We use partner symmetries for lifting invariant solutions
of elliptic and hyperbolic $CMA$, that satisfy equations of lower
dimensions, to non-invariant solutions of $CMA$.

A symmetry reduction of a partial differential equation reduces by
one the number of independent variables in the original equation,
so that the reduced equation is easier to solve. Its solutions are
solutions of the original PDE that are invariant under the
symmetry that was used in the reduction. Even if we found
non-invariant solutions of the reduced equation, it would only
mean that no further symmetry reduction was made and they would
still be invariant solutions of the original equation.

For complex Monge-Amp\`ere equations, we have shown that partner
symmetries provide a procedure reverse to the symmetry reduction:
a lift of invariant solutions of $CMA$ to non-invariant solutions
of $CMA$. This means holographic property of the symmetry used for
the reduction, i.e. the information on solutions is not completely
lost under the reduction but can be reconstructed for a certain
class of non-invariant solutions.

We have performed such a procedure for the elliptic and hyperbolic
$CMA$ and obtained non-invariant solutions of these equations.
Using these solutions as metric potentials, it is possible to
obtain gravitational metrics of Euclidean and ultra-hyperbolic
signatures that have no Killing vectors \cite{lift}.

We are now in the process of developing a modified procedure of
the lift from non-invariant solutions of the elliptic Boyer-Finley
equation to non-invariant solutions of the elliptic $CMA$.

\section*{Acknowledgements}

Authors thank Y. Nutku for useful discussions. The research of MBS
is partly supported by the research grant from Bogazici University
Scientific Research Fund, research project No. 07B301.

\end{document}